\documentstyle[11pt,newpasp,twoside,epsf]{article}
\def\edcomment#1{\iffalse\marginpar{\raggedright\sl#1\/}\else\relax\fi}
\reversemarginpar

\begin{document}

\title{The Evolution of Disks and Winds in Dwarf Novae Outbursts}

\author{C.\ S.\ Froning, K.\ S.\ Long} 

\affil{Space Telescope Science Institute, 3700 San Martin Drive, Baltimore,
MD 21218 USA}

\author{J.\ E.\ Drew}
\affil{Imperial College of Science, Technology and Medicine, Blackett 
Laboratory, Prince Consort Rd., London SW7 2BZ UK}

\author{C.\ Knigge}
\affil{Dept.\ of Physics and Astronomy, University of Southampton, 
Southampton SO17 1BJ UK}

\author{D.\ Proga}
\affil{Laboratory of High Energy Astrophysics, NASA Goddard Space Flight 
Center, Greenbelt, MD 20771 USA}

\author{J.\ A.\ Mattei}
\affil{American Association of Variable Star Observers, 25 Birch St., 
Cambridge, MA 02138 USA}

\begin{abstract}
Far ultraviolet (FUV) observations are excellent probes of the inner
accretion disk, disk outflows, and the mass-accumulating white dwarf
in cataclysmic variables.  Here we study the contrasting behavior of
two canonical dwarf novae in outburst by presenting FUSE FUV (904 --
1187~\AA) spectroscopy of U~Gem and SS~Cyg.  We observed each system
four times during a single outburst.  The outburst peak and early
decline spectra of SS~Cyg are well fit by models of a steady-state
accretion disk and a biconical wind.  A broad, blueshifted OVI
wind-formed absorption line is the only strong spectral feature.  In
late outburst decline, O~VI and C~III lines are seen as broad emission
features and the continuum has flattened.  In U~Gem, the continua of
the optical outburst plateau spectra are plausibly fit by accretion
disk model spectra.  The spectra also show numerous narrow,
low-velocity absorption lines that do not originate in the inner
accretion disk.  We discuss the line spectra in the context of partial
absorption of the FUV continuum by low-velocity, vertically-extended
material located at large disk radii.  The late outburst decline
spectrum of U~Gem is dominated by the white dwarf.  WD model fits
confirm the sub-solar C and super-solar N abundances found in earlier
studies.
\end{abstract}

\section{Introduction}

The outburst cycles of dwarf novae (DN) allow observers to probe the
evolution of disk-accreting systems as they undergo quasi-periodic
perturbations.  During outburst, changes in the structure of the disk
as a function of mass accretion rate, the link between the state of
the disk and outflows, and the impact of the accreted material on the
structure and properties of the mass-accreting primary can be
examined.  The far-ultraviolet (FUV) is well-suited for such
studies. The FUV continuum, where the hot disk peaks in flux, is
sensitive to the structure of the inner disk.  The FUV is also rich in
spectral lines that trace the temperature and ionization structure of
disk, winds, and the white dwarf/boundary layer region in DN.  In this
light, we have pursued FUV spectroscopy of two canonical DN, U~Gem and
SS~Cyg, during the peaks and declines of their outbursts.

\section{The FUSE Observations}

We observed U~Gem during its 2000 March outburst.  Three of the
observations were obtained on optical outburst plateau, the fourth
during late outburst decline, about two days before the return to
optical quiescence.  We observed SS~Cyg during the 2000 November
narrow outburst.  The first observation was obtained at outburst peak
and the following three during the decline, with the last observation
occuring on the final day of the optical outburst.  All observations
were obtained with FUSE, which covers the 904 -- 1187~\AA\ wavelength
at a spectral resolution through its large aperture of
R$\sim$12,000. We obtained multiple spectra spread over several binary
orbits during each observation.  The typical exposure times of
individual spectra were $\sim$500~s.  We combined the spectra of each
observation to create time-averaged spectra at a resolution of
$\sim$0.1~\AA.  Further details of the data reduction and processing
can be found in Froning et al.\ (2001).

\section{SS Cygni at Outburst Peak and in Decline}

\begin{figure}
\plotone{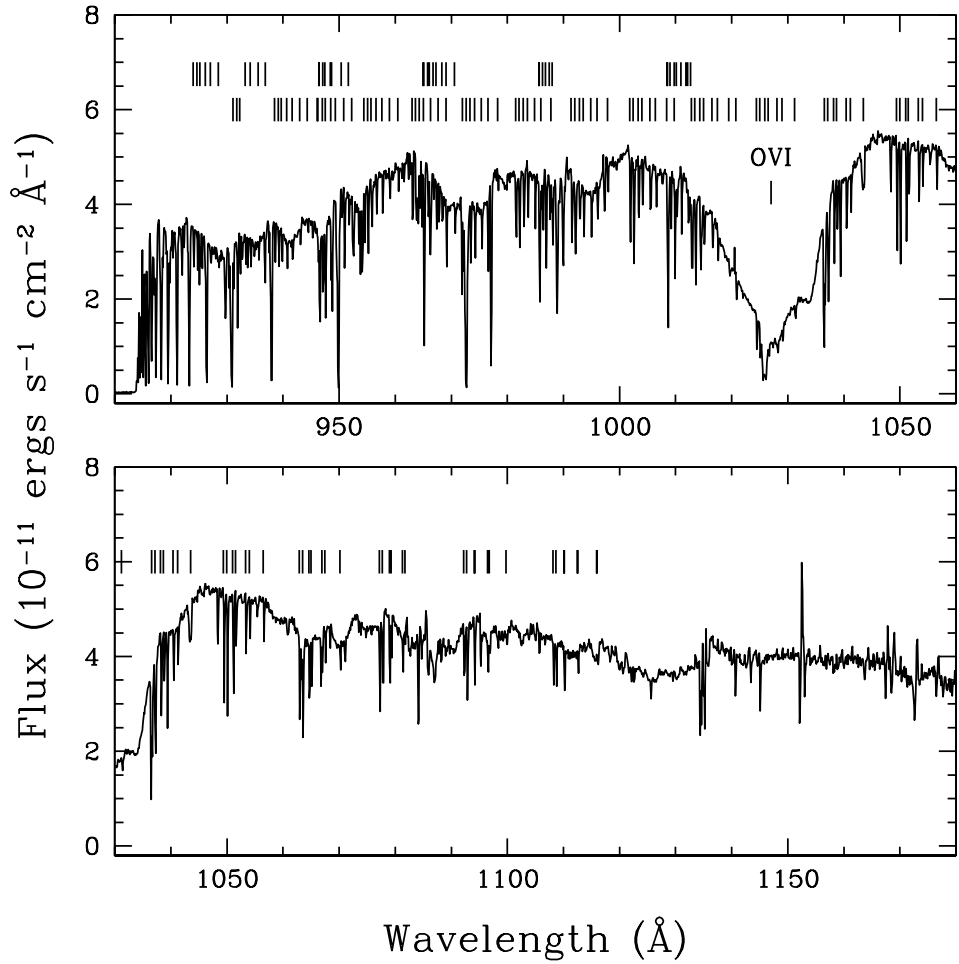}
\caption{The Obs.\ 1 time-averaged spectrum of SS~Cyg, acquired at the
peak of the 2000 November outburst. The total observation time was
5600~sec. Identified Werner and Lyman band transitions of molecular
hydrogen are marked above the spectrum.  The O~VI absorption line is
also labeled. }
\end{figure}

The time-averaged spectrum of the first observation of SS~Cyg,
obtained at outburst peak, is shown in Fig.\ 1.  The spectrum is
dominated by interstellar lines of H~I, molecular hydrogen, and
metals.  The interstellar H$_{2}$ Werner and Lyman band transitions
are indicated in Fig.\ 1.  The only prominent feature from SS~Cyg in
the spectrum is a broad trough at the location of the Ly$\alpha$ and
O~VI $\lambda\lambda$1032,1038~\AA\ lines.

\begin{figure}
\plotone{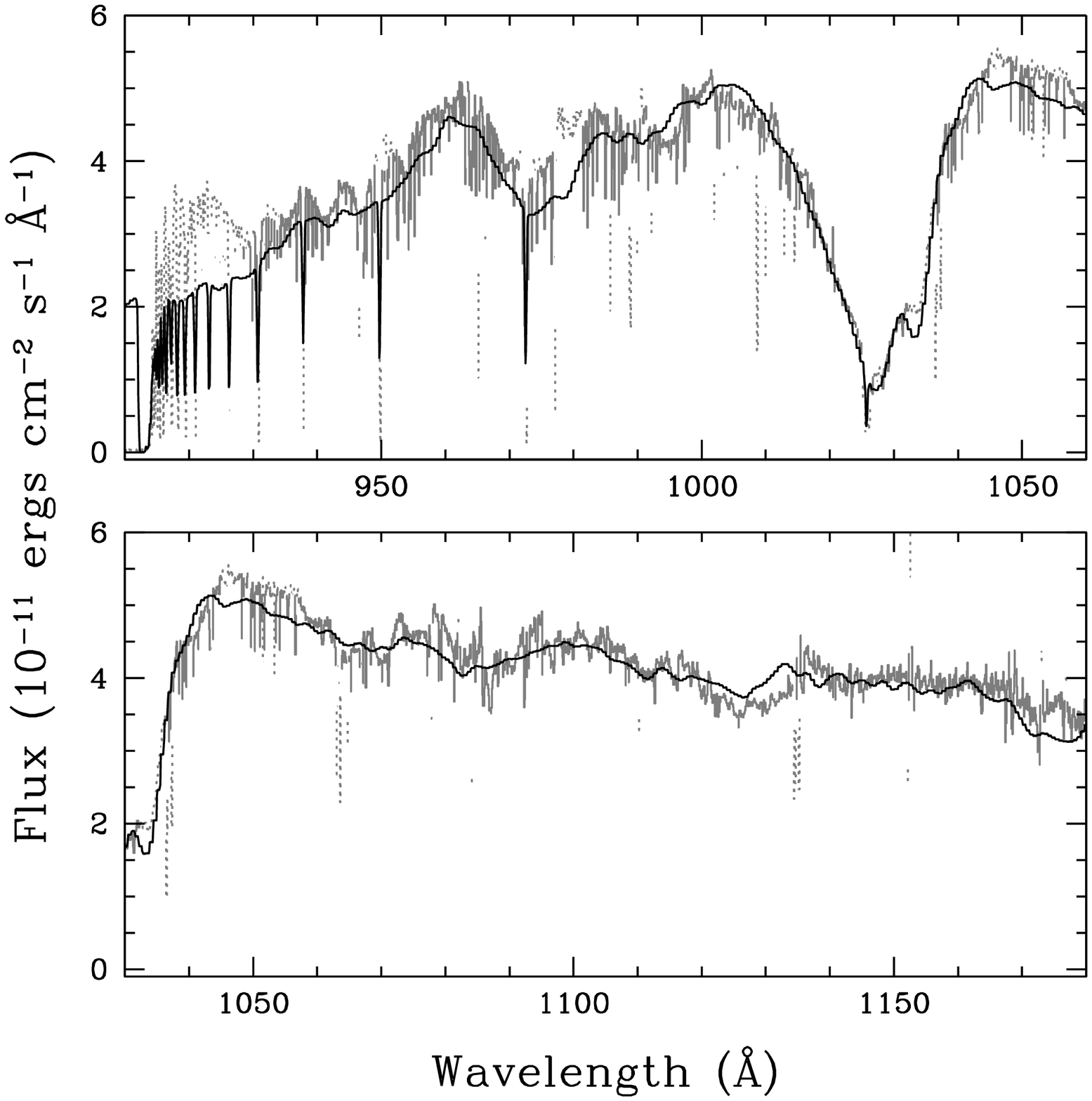}
\caption{The spectrum of Figure 1 with a steady-state accretion disk
plus a rotating, biconical wind model superimposed. The model was fit
to the spectrum in two iterations; the parts of the spectrum masked
out in the second iteration are plotted with dotted lines.}
\end{figure}

We fit accretion disk model spectra to the spectrum shown in Fig.\ 1.
The model spectra are constructed from weighted, summed stellar
atmosphere spectra at the appropriate temperature and gravity for each
disk annulus assuming a steady-state disk.  The model atmosphere and
spectral synthesis codes TLUSTY and SYNSPEC were used to construct the
stellar spectra (Hubeny 1988; Hubeny, Lanz \& Jeffery 1994).  To the
disk models, we added a rotating, biconical wind model using the Monte
Carlo radiative transfer code, Python (Long \& Knigge, these
proceedings).  A representative --- not to be considered unique ---
fit to the spectrum at outburst peak is shown in Fig.\ 2.  The mass
accretion rate in the disk for the model shown is
$4.4\times10^{-9}$~M$_{\odot}$~yr$^{-1}$ and the mass loss rate in the
wind is $4.0\times10^{-11}$~M$_{\odot}$~yr$^{-1}$.  The model is a
good qualitative fit to the shape of the spectrum, although it
under-predicts the flux for wavelengths $<$930~\AA.  It is unclear
whether the deviation of model from data near the Lyman limit is
caused by an additional flux source not represented in the model or by
some missing element in the model physics.

The second observation, obtained five days after the first, has a
similar spectrum, although the FUV flux dropped by 60\% (at 1000~\AA)
between the observations.  By the time the third observation was
obtained, 8 days after the first and 10 days after outburst start, the
FUV flux at 1000~\AA\ was
$3.8\times10^{-12}$~ergs~cm$^{-2}$~s$^{-1}$~\AA$^{-1}$, down from
$5.0\times10^{-11}$~ergs~cm$^{-2}$~s$^{-1}$~\AA$^{-1}$ in the first
observation and $1.8\times10^{-11}$~ergs~cm$^{-2}$~s$^{-1}$~\AA$^{-1}$
in the second.  In contrast to the earlier spectra, the FUV continuum
is fairly flat in the third observation. Although still in absorption
at blue wavelengths, the O~VI line developed a sharp, red emission
peak by the third observation.  In addition, C~III $\lambda$977~\AA,
N~III $\lambda$990~\AA, and C~III $\lambda$1176~\AA\ appear strongly
in emission.

In the final observation, obtained 13 days after outburst start and on
the last day of the optical outburst, the FUV flux had declined to
$5.0\times10^{-13}$~ergs~cm$^{-2}$~s$^{-1}$ \AA$^{-1}$ (again at
1000~\AA).  The shape of the spectrum is similar to a HUT spectrum of
SS~Cyg in quiescence, although the 1000~\AA\ flux is 40\% higher in
the FUSE spectrum.  The dominant features are the C~III, N~III, and
O~VI emission lines, the latter now completely in emission. The FWHM
of the features, 3600~km~s$^{-1}$ for the O~VI doublet (assuming no
Ly$\beta$ contribution to the feature) and 2500~km~s$^{-1}$ for the
C~III $\lambda$1176~\AA\ blend, are similar to those measured in the
HUT spectrum, but the EWs are at least twice as large, and the line
fluxes 4 times as large, as those in the quiescent HUT spectrum.

\section{U Geminorum}

Additional analysis of the FUSE FUV spectroscopy of U~Gem in outburst
can be found in Froning et al.\ (2001).

\subsection{The Outburst Plateau}

\begin{figure}
\plotone{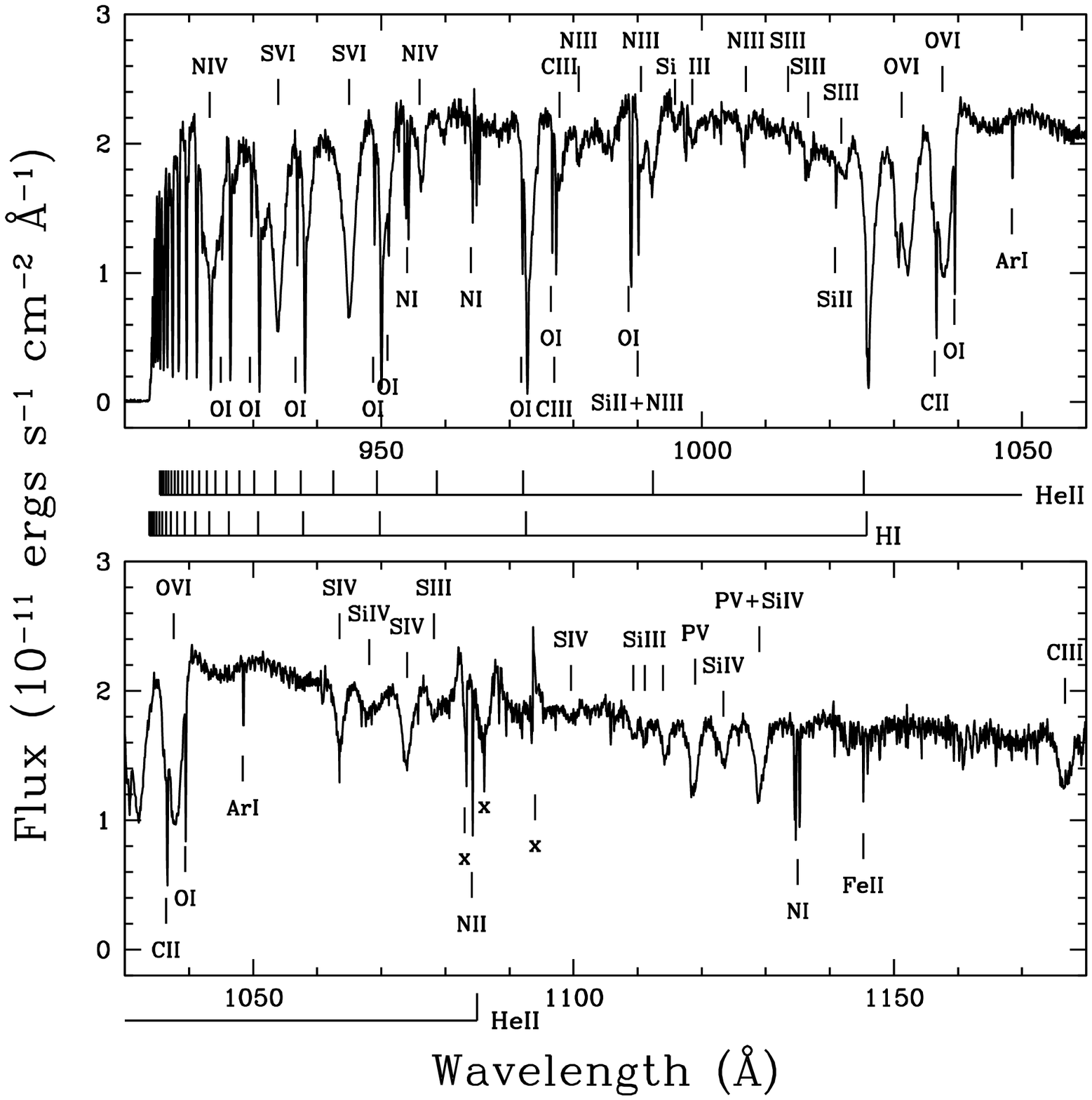}
\caption{The Obs.\ 1 time-averaged spectrum of U~Gem, acquired at the
peak of the 2000 March outburst.  The locations of HI and HeII lines
are labeled below the spectrum.  Absorption lines of metals intrinsic
to U~Gem are labeled above the spectrum, while prominent interstellar
metal lines are labeled below the spectrum.  Instrumental features are
marked with an ``x''. }
\end{figure}

The first three observations of U~Gem took place during the optical
plateau phase of the outburst.  The FUV continuum declined by 10\%
between the first and second observations and by 12\% between the
second and third observations.  The continuum decline was gray.  The
time-averaged spectrum of the first observation is shown in Fig.\ 3. In
contrast to the spectrum of SS~Cyg, U~Gem's FUV spectrum is rich in
features.  The spectrum has narrow lines from interstellar absorption,
but there are also numerous absorption lines from U~Gem, including
transitions of H~I, He~II, and metals in two to five times ionized
states.  With the exception of a weak bump to the red of O~VI, there
are no emission features in any of the spectra.

We fit accretion disk model spectra to the time-averaged spectrum of
each plateau observation using the models presented in Section 3.  We
did not include a wind in the model because there is no sign of a
strong wind in the FUV in U~Gem (see below).  Our best-fit model to
the spectrum of the first observation has a mass accretion rate in the
disk of $7\times10^{-9}$~M$_{\odot}$~yr$^{-1}$, close to values found
for U~Gem in previous outbursts (Panek \& Holm 1984; Sion et
al. 1997).  Since the shape of the FUV spectrum didn't change from
observation to observation, the temperature and mass accretion rate in
the disk remained steady while U~Gem was on optical outburst plateau,
although the effective emitting area of the disk dropped by $\sim$22\%
over the three observations.  The FUV flux dropped before the optical
flux did, indicating that the outburst decline was inside-out.

The accretion disk model provides a plausible fit to the shape of the
FUV continuum in the three observations, although as with SS~Cyg, the
model under-predicts the observed flux at short wavelengths
($<$960~\AA).  The boundary layer is known to be bright in U~Gem in
outburst from EUVE observations (Long et al.\ 1996) and may contribute
to the FUV flux. Using parameters for the boundary layer from Long et
al., we estimate that a 138,000~K blackbody with a size comparable to
that of the WD would contribute $\sim$25\% of the observed flux shown
in Fig.\ 1 at 915~\AA\ but only $\sim$13\% at 1180~\AA, indicating
that boundary layer emission may contribute to the excess observed
flux near the Lyman limit.

The absorption lines in the outburst plateau spectra are narrow, with
FWHM ranging from 250 -- 850~km~s$^{-1}$.  They are also at low
velocity, and while they move in phase with the WD over the binary
orbit, at no time are they more than 700~km~s$^{-1}$ from their rest
positions.  Most of the absorption lines have smooth profiles and the
line shapes and EWs at a given orbital phase are the same in all three
observations.  The O~VI and S~VI doublet flux ratios indicate that
these lines are optically thick, but none of the absorption lines are
dark in their line centers.

The O~VI line behaves differently from the other lines in the
spectrum.  As noted above, the weak bump to the red of the doublet
absorption is the only hint of emission in the FUV and may indicate
the presence of a weak wind in U~Gem in outburst, seen only in the
most energetic FUV transition.  This supposition is supported by the
presence of narrow (FWHM $\sim$ 100~km~s$^{-1}$), blue-shifted
($-500$~km~s$^{-1}$) dips that appear sporadically in the O~VI lines.
These dips have also been seen in wind-formed lines of other CVs and
X-ray binaries, although their origin is not yet understood (see,
e.g., Hartley, Drew, \& Long, these proceedings).  Aside from this
weak signature in O~VI, the FUV absorption lines in U~Gem are too
narrow and insufficiently blueshifted to originate in a standard CV
outflow.  Wind lines are seen in the EUV spectrum of U~Gem in outburst
(Long et al.\ 1996), suggesting that the bulk of the wind may be
highly ionized and mostly invisible in the FUV.

If the FUV lines (aside from O~VI) do not originate in a standard
wind, another source must be sought. The inner accretion disk, source
of the FUV continuum, is not plausible, as the inner disk velocities
--- several thousand km~s$^{-1}$ --- are much too large to be
reconciled with the small line widths.  The FUV absorption lines must
originate at large disk radii if the absorbing region moves at
Keplerian velocities.  An outer disk origin is supported by the
presence of orbital variability in the FUV lines.  Between orbital
phases 0.5 -- 0.8 in all three plateau observations, the central
depths of the absorption lines increase and several low ionization
transitions of He~II, C~III, N~III, Si~III, and S~III not seen at
other phases appear.  The increase in line absorption occurs at the
same orbital phases as X-ray and EUV light curve dips observed in
U~Gem (Mason et al.\ 1988; Long et al.\ 1996; Szkody et al.\ 1996).
The light curve dips are due to absorption of the central source by
vertically extended material located at large disk radii.  This
material is believed to result from the interaction between the mass
accretion stream and the accretion disk, either as a disk bulge or by
the stream overflowing the disk.

The increase in the FUV line absorption occurs at the same orbital
phases as the X-ray and EUV light curve dips, so it is reasonable to
conclude that the same absorbing region is responsible for both
effects.  In that case, the FUV line spectrum in U~Gem in outburst is
caused by absorption by a vertically extended disk chromosphere with
enhanced absorption at phases 0.5 -- 0.8 from extra material tied to a
disk bulge or stream overflow.  A simple model for U~Gem of a
Keplerian, pure absorption chromosphere shows that material located
from 20 -- 60 WD radii with a scale height of $2\times10^{10}$~cm can
match the velocities and depths of the FUV absorption lines.

\subsection{Late Outburst Decline}

\begin{figure}
\plotone{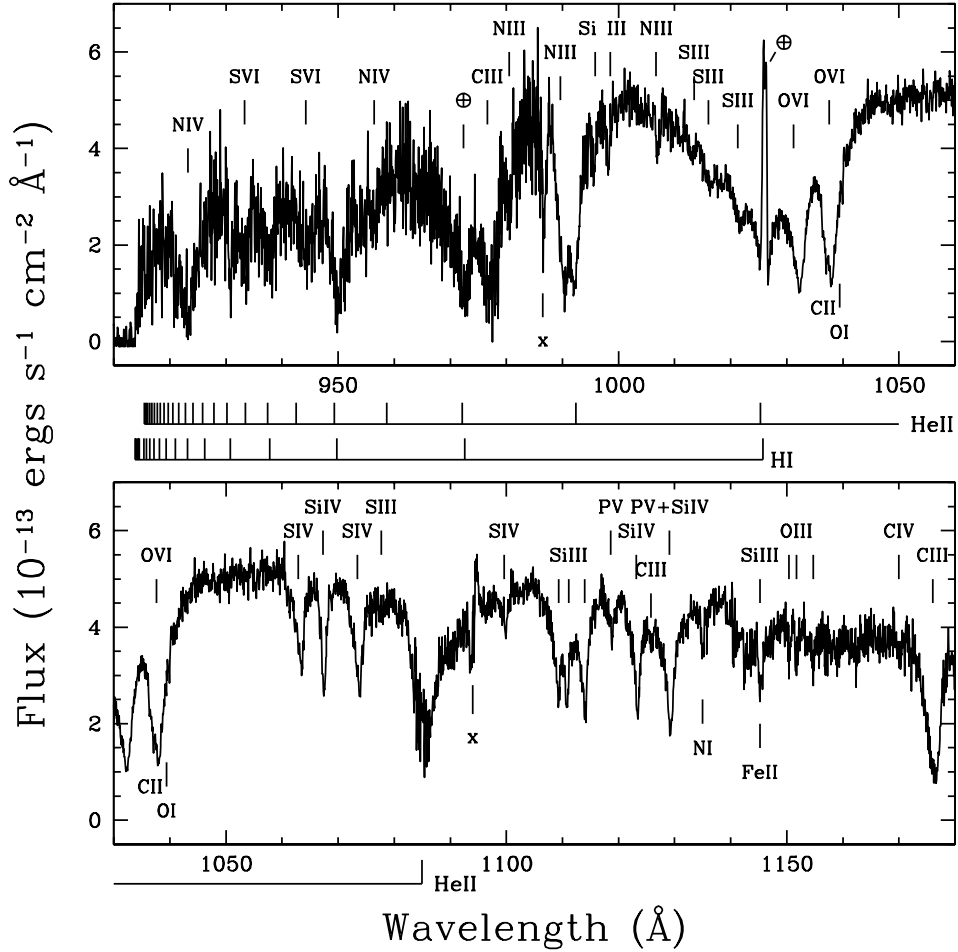}
\caption{The Obs.\ 4 time-averaged spectrum of U~Gem, acquired when
the system was at m$_{V} \sim$ 12, about 2 days before the return to
optical quiescence.  The circled crosses indicate lines of terrestrial
airglow; other features are labeled as in Figure 3. The individual
(300 sec exposure time) spectra comprising the time-averaged spectrum
were shifted in velocity to remove the orbital motion of the white
dwarf before combining.}
\end{figure}

The final observation of U~Gem was obtained two days before the return
to optical quiescence.  The time-averaged spectrum is shown in Fig.\
4.  The FUV continuum flux is a factor of 30 below that of the
previous observation.  The spectrum is very similar in shape to a HUT
spectrum of U~Gem obtained during early quiescence, which was well fit
by white dwarf (WD) model spectra (Long et al.\ 1993).  The similarity
of the FUSE spectrum to the quiescent spectrum suggests that although
the visible outburst was not over when the final FUSE observation took
place, the accretion disk had fully faded in the FUV, exposing the WD.
We fit model WD spectra to the time-averaged spectrum using TLUSTY and
SYNSPEC.  Our best fit model of a single temperature WD has a
temperature T$_{WD}$ = 43,410~K, a gravity $\log g$ = 8.0, and a WD
rotation rate of 200~km~s$^{-1}$.  The model is a good qualitative fit
to the spectrum.  Most of the lines present in the data are also
present in the model, and the shapes of the continuum and the lines
are reasonably well reproduced by the model.

In quiescent observations of U~Gem, the UV flux declines between
outbursts (30\% at 1400~\AA; Kiplinger, Sion, \& Szkody 1991; Long et
al.\ 1994).  WD models of the FUV spectra have shown that only a
portion of the WD can cool during quiescence if the observed flux
decline and the change in the WD temperature in the models are to be
reconciled (Long et al.\ 1993).  The FUV flux (at 1000~\AA) in the
FUSE late outburst spectrum is 30\% higher again than the flux
observed in early quiescence. A second temperature component on the WD
is not required to reconcile the models to the observed fluxes in this
case, as the 5000~K difference between the WD temperatures in the late
outburst and the early quiescent models is consistent with the
observed flux difference of 30\% between those observations.  The
addition of a second temperature component to the model of the FUSE
spectrum provides a slight statistical and qualitative improvement to
the fit, however.

Models of the quiescent spectrum of U~Gem have also indicated WD metal
abundances that differ from solar; in particular, C is significantly
sub-solar and N is super-solar, suggesting the presence of material on
the WD surface that has undergone CNO processing (Sion et al.\ 1998,
Long \& Gilliland 1999).  The FUSE spectra cover a different
wavelength range than the previously modeled (HST) spectra and provide
an independent check on the surface abundances of the WD.  The WD
models of the FUSE spectrum confirm the non-solar abundances on the WD
in U~Gem.  The FUV spectrum contains a number of C lines that are
significantly over-predicted by WD models at solar abundances and are
better fit when the C abundance is of order 0.1 solar.  Similarly, the
numerous N lines in the FUSE spectrum are under-predicted by solar
abundance WD models and are better fit when the N abundance is
increased to several times solar.  Fits to the other metal lines in
the FUSE spectrum indicate no substantial deviations from solar.

\acknowledgements{Based on observations with the NASA-CNES-CSA Far
Ultraviolet Spectroscopic Explorer.  FUSE is operated for NASA by the
Johns Hopkins University under NASA contract NAS5-32985.  We
gratefully acknowledge the financial support of NASA through grant
NAG5-9283.  We wish to thank the AAVSO and its observers for notifying
us of the outbursts of SS~Cyg and U~Gem and monitoring the progress of
the outbursts. We also thank the FUSE staff for their heroic efforts
in scheduling and conducting the observations.}


\begin{references}

\reference Froning, C.\ S., Long, K.\ S., Drew, J.\ E., Knigge, C., \& 
Proga, D. 2001, ApJ, accepted

\reference Kiplinger, A.\ L., Sion, E.\ M., \& Szkody, P. 1991, \apj,
366, 569

\reference Long, K.\ S.\ \& Knigge, C. 2001, these proceedings

\reference Long, K.\ S.\ \& Gilliland, R.\ L. 1999, \apj, 511, 916

\reference Long, K.\ S., Mauche, C.\ W., Raymond, J.\ C., Szkody, P., \& 
Mattei, J.\ A. 1996, \apj, 469, 841

\reference Long, K.\ S., Sion, E.\ M., Huang, M., \& Szkody, P. 1994,
\apj, 424, L49

\reference Long, K.\ S., Blair, W.\ P., Bowers, C.\ W., Davidsen, A.\ F.,
Kriss, G.\ A., Sion, E.\ M., \& Hubeny, I. 1993, \apj, 405, 427

\reference Hartley, L., Drew, J.\ E., \& Long, K.\ S. 2001, these
proceedings

\reference Hubeny, I., Lanz, T., \& Jeffery, C.~S. 1994, Newsletter on
Analysis of Astronomical Spectra (St.\ Andrews: St.\ Andrews Univ.), 20,
30
\reference Hubeny, I. 1988, Comput. Phys. Comm., 52, 103

\reference Mason, K.\ O., C\'{o}rdova, F.\ A., Watson, M.\ G., \& 
King, A.\ R. 1988, \mnras, 232, 779

\reference Panek, R.\ J.\ \& Holm, A.\ V. 1984, \apj, 277, 700

\reference Sion, E.\ M., Cheng, F.\ H., Szkody, P., Sparks, W.,
Gaensicke, B., Huang, M., \& Mattei, J. 1998, \apj, 496, 449

\reference Sion, E.\ M., Cheng, F.\ H., Szkody, P., Huang, M., Provencal,
J., Sparks, W., Abbott, B., Hubeny, I., Mattei, J., \& Shipman, H. 
1997, \apj, 483, 907

\reference Szkody, P., Long, K.\ S., Sion, E.\ M., \& Raymond, J.\
C. 1996, ApJ, 469, 834

\end{references}
\end{document}